\documentclass[fleqn,usenatbib]{mnras}

\usepackage[T1]{fontenc}

\usepackage{newtxtext,newtxmath}
\usepackage{subfigure}

\usepackage{graphicx}	% Including figure files
\usepackage{amsmath}	% Advanced maths commands

\usepackage{float}
\usepackage{arydshln}
\title[Exploring the properties of COS-87259]
{Exploring the properties of the obscured hyperluminous quasar COS-87259 at \boldmath$z=6.853$}

\author[Charalambia Varnava]{\parbox{\linewidth}{Charalambia Varnava$^{1}$\thanks{E-mail: \href{mailto:varnava.haris@gmail.com}{varnava.haris@gmail.com} (CV)}, 
Andreas Efstathiou$^{1}$ and Duncan Farrah$^{2,3}$
}
\\
\\
$^{1}$School of Sciences, European University Cyprus, Diogenes street, Engomi, 1516 Nicosia, Cyprus\\
$^{2}$Department of Physics and Astronomy, University of Hawaii, 2505 Correa Road, Honolulu, HI 96822, USA\\
$^{3}$Institute for Astronomy, University of Hawaii, 2680 Woodlawn Drive, Honolulu, HI 96822, USA
}

\date{Accepted 2024 September 23; Received 2024 September 9; in original form 2024 August 9}

\pubyear{2024}

\begin{document}
\label{firstpage}
\pagerange{\pageref{firstpage}--\pageref{lastpage}} 
\maketitle

\begin{abstract}
In this paper we explore the properties of the $z=6.853$ obscured hyperluminous quasar COS-87259, discovered in the Cosmological Evolution Survey (COSMOS) field, with our recently developed Bayesian spectral energy distribution (SED) fitting code SMART (Spectral energy distributions Markov chain Analysis with Radiative Transfer models). SMART fits SEDs exclusively with multicomponent radiative transfer models that constitute four different types of pre-computed libraries for the active galactic nucleus (AGN) torus, the starburst and the spheroidal or disc host. We explore two smooth radiative transfer models for the AGN torus and two two-phase models, in order to put constraints on the AGN fraction of the galaxy, the black hole mass and its star formation rate (SFR). We find that either of the smooth tapered disc or the two-phase flared disc models provide a good fit to the SED of COS-87259. The best-fitting models predict an AGN fraction in the range $86-92$ per cent, a bolometric AGN luminosity of $5.8-10.3 \times 10^{13} L_\odot$, a black hole mass of $1.8-3.2 \times 10^{9} M_\odot$ (assuming the quasar is accreting at the Eddington limit) and an SFR in the range $1985-2001~M_\odot yr^{-1}$. The predicted space density of such objects in the redshift range $4-7$ is $\sim 20$ times more than that of co-eval unobscured quasars. 
\end{abstract}


\section{Introduction}\label{sec:intro}

With the advent of \textit{JWST}, quasars have now been observed out to $z>10$. \cite{bog24}, for example, reported the detection of an X-ray luminous active galactic nucleus (AGN) in a gravitationally lensed galaxy identified by \textit{JWST} behind the cluster lens Abell 2744 at  $z\sim10.3$. The UNCOVER project, carried out with \textit{JWST} recently, reported 26 candidate red AGNs at $3<z<7$ behind the same lensing cluster. None of these red and potentially obscured AGNs have been detected by Atacama Large Millimeter/submillimeter Array (ALMA). \cite{bosm24} reported the first rest frame infrared spectrum of the $z>7$ quasar J1120+0641, obtained with the MIRI-MRS instrument onboard \textit{JWST}.

All of these high redshift quasars, however, currently lack data in the rest frame mid-infrared to submillimetre range, to constrain their star formation rate (SFR), AGN torus luminosity and black hole mass. Programs to observe these high $z$ quasars with other facilities are underway. \cite{endsley23} confirmed with ALMA observations the redshift of the hyperluminous infrared galaxy COS-87259 at $z=6.853$, which they discovered in the Cosmological Evolution Survey (COSMOS) field. Being in the best studied extragalactic field, COS-87259 has a spectral energy distribution (SED) that is well sampled in the rest frame far-infrared, as it has detections by Herschel, SCUBA-2, as well as ALMA. Although the bolometric luminosity of the galaxy appears to be dominated by an AGN, the galaxy is predicted to have an SFR of $2439.4^{287.9}_{-287.9} M_\odot yr^{-1}$ and a stellar mass $25.8^{10.6}_{-10.6} \times 10^{10} M_\odot$ according to \cite{endsley23}\footnote{All estimates of stellar mass and SFR in this paper either assume or have been converted to a Salpeter initial mass function.}, who fitted the SED of COS-87259 with an SED fitting code that features (semi-)empirical templates of AGN and star-forming dust emission \citep{lyu22}. \cite{vanm23} estimated the stellar mass of COS-87259 with a number of other SED fitting codes and concluded that its stellar mass may not be so extreme. However, most of the codes explored by \cite{vanm23} did not consider the infrared to submillimetre emission or the contribution from an AGN, so can not provide alternative estimates of the SFR and AGN torus luminosity.  The fact that COS-87259 was discovered in the 1.7 deg$^2$ COSMOS field is very significant. As discussed by \cite{endsley23}, the implied space density of obscured quasars at $z>6$ appears to be about three orders of magnitude higher than that of coeval quasars of the same luminosity \citep{wang19}. This is unlikely to be due to magnification of the source with an intervening galaxy. \cite{endsley23} estimate that COS-87259 may be slightly magnified, with a magnification factor in the range $1.15-1.25$. However, for easier comparison with earlier work, we follow \cite{endsley23} and do not apply any magnification corrections in the physical quantities we report in this paper. We also note that we assume the same cosmology as \cite{endsley23}. 

There are a number of AGN torus models in the literature, as well as a number of SED fitting codes. It is still an open question how well we can determine the SFR and AGN torus luminosity of galaxies that host obscured quasars at $z>4$. We aim to quantify these uncertainties in this paper, using our recently developed Bayesian SED fitting code SMART (Spectral energy distributions Markov chain Analysis with Radiative Transfer models; \citealt{smart,ascl}). SMART can fit an SED with eight different combinations of radiative transfer models, each employing a different AGN torus model \citep{efstathiou95,fritz06,sieb15,stal16}, in addition to a starburst and a host galaxy model, which can be assumed to be either spheroidal or disc. The method can also include a component of polar dust associated with the AGN \citep{efstathiou06}, but due to the limited data we don't include this component in the analysis of this paper. Our method allows us to determine the most reliable estimates of the AGN fraction and SFR of COS-87259 that can be obtained with the currently available data. Our analysis builds on the results of \cite{efs21}, who discovered HELP\_J100156.75+022344.7, a similar object to COS-87259 at $z\sim4.3$, also in the COSMOS field, and fitted it with the Bayesian SED fitting code SATMC (Spectral energy distribution Analysis Through Markov Chains; \citealt{john13}). \cite{smart} carried out a similar analysis of HELP\_J100156.75+022344.7 with the four combinations of models discussed in this paper and confirmed the results of \cite{efs21}.

This paper is organized as follows: In Section \ref{sec:data} we decribe the data, in Section \ref{sec:description} we describe the models and the SED fitting method and in Section \ref{sec:results} we present and discuss our results. In Section \ref{sec:conclusion} we present our conclusions. Throughout this work we assume $H_0=70$\,km\,s$^{-1}$\,Mpc$^{-1}$, $\Omega=1$ and $\Omega_{\Lambda}=0.7$.

\section{Data}\label{sec:data}

COS-87259 was identified by \cite{endsley22} as one of 41 UV-bright (M$_{UV} \lessapprox -21.25$) $z \simeq  6.6-6.9$ Lyman-break galaxy candidates in the COSMOS field. The source is undetected at all wavelengths shorter than 9600 \AA, as expected from a Lyman-break galaxy at $z \sim 6.8$. The data we used in this work are tabulated in Table \ref{tab:data}. We have a total of 18 photometry points.

The [C$_{\rm II}$] and far-infrared dust emission in the ALMA image of \cite{endsley23} consists of a compact component and a more extended component. The flux density from the two components is $1.48 \pm 0.05$ and $1.23  \pm 0.15$ mJy, respectively, yielding a total 158 $\mu m$ dust continuum flux density of $2.71 \pm 0.16$ mJy for COS-87259.
The physical sizes of the compact and more extended components are ($1.4  \pm 0.2$ kpc)  $\times$ ($1.0 \pm 0.1$ kpc) and ($7.8 \pm 1.0$ kpc) $\times$ ($3.4 \pm 0.5$ kpc), respectively.

\begin{table}
	\centering
\caption{Photometry of COS-87259 used in this work. The data are taken from Table 1 of Endsley et al. (2022), except for the ALMA point which is from Endsley et al. (2023).}
	\label{tab:data}
	\begin{tabular}{llll} % two columns, alignment for each
		\hline
		Wavelength  &  Flux  &  Error &  Instrument \\
         $\mu m$    &  $\mu$Jy &  $\mu$Jy &         \\
		\hline 
0.973 &  0.20 &  0.06  & HSC nb973 \\
1.   & 0.18  & 0.04    &   HSC y \\
1.02  & 0.22 &  0.04   & VIRCam Y \\
1.25  & 0.37 &  0.05   &  VIRCam J \\
1.6  &  0.44 &  0.12   & WFC3 F160W \\
1.6  &  0.56 &  0.05  & VIRCam H \\
2.1  &  0.67 &  0.09  & VIRCam Ks \\

3.6  &  2.75 &  0.10  & Spitzer/IRAC 3.6 $\mu m$ \\
4.5  &  2.77 &  0.12  & Spitzer/IRAC 4.5 $\mu m$ \\
5.8  &  4.42 &  1.04  & Spitzer/IRAC 5.8 $\mu m$\\

24  &  179.7 &  6.4  & Spitzer/MIPS 24 $\mu m$ \\
100 &   5100 &  1370 & Herschel/PACS 100 $\mu m$ \\
160 &  10130 &  4260 & Herschel/PACS 160 $\mu m$\\
250 &  8830  &  1120  & Herschel/SPIRE 250 $\mu m$ \\
350 &  8090  &  1600  & Herschel/SPIRE 350 $\mu m$\\
500 &  11130 &  2220 &  Herschel/SPIRE 500 $\mu m$\\
850 &  6300  &  1650  & SCUBA-2 850 $\mu m$ \\

1239 & 2710   & 160  & ALMA  242 GHz\\
		\hline
	\end{tabular}
\end{table}

\section{Description of the models and SED fitting method}\label{sec:description}

Our method, described in more detail in \cite{smart}, allows us to explore the impact of four different AGN torus models and therefore constrain the properties of the obscuring torus, but also quantify the uncertainties in the AGN fraction, black hole mass and SFR of the fitted galaxy. Each torus model is fitted in combination with the starburst model of \cite{efstathiou00} as updated by \cite{efstathiou09} and the spheroidal host model of \cite{efs21}:

\begin{enumerate}

\item The smooth AGN torus model originally developed by \cite{efstathiou95}, which is part of the CYGNUS (CYprus models for Galaxies and their NUclear Spectra) collection of radiative transfer models. More details of the implementation of this combination of models within MCMC codes are given in \cite{efs21}, \cite{efs22} and \cite{smart}. This model assumes a tapered disc geometry (the thickness of the disc increases linearly with distance from the black hole in the inner part of the torus, but assumes a constant thickness in the outer part).

\item The smooth AGN torus model of \cite{fritz06}, which assumes a flared disc geometry (the thickness of the disc increases linearly with distance from the black hole).

\item The two-phase AGN torus model SKIRTOR of \cite{stal16}, which also assumes a flared disc geometry.

\item The two-phase AGN torus model of \cite{sieb15}, which assumes that dust covers the whole sphere around the black hole, i.e. the half-opening angle of the torus is assumed to be zero. Also, unlike the other three torus models listed above, this model assumes that dust grains are fluffy and therefore have a higher emissivity in the far-infrared and submillimetre. This model is therefore expected to give the stronger contribution from the AGN in that part of the spectrum. 
\end{enumerate}

Each AGN torus model has four free parameters. The SKIRTOR and \cite{fritz06} models have two additional parameters which we fix. As argued in \cite{smart}, fixing the parameter $p$ of the SKIRTOR model to the value of 1, gives the best agreement for the silicate absorption features of obscured quasars. The parameter $q$ in SKIRTOR concerns the azimuthal dependence of the density distribution and it is unlikely to have much impact on the silicate features. As in \cite{smart}, we have fixed this value at 1. We have similarly fixed two of the parameters of the \cite{fritz06} model, as detailed in Table \ref{tab:parameters}.

The starburst model has three parameters, which are the initial optical depth of the molecular clouds that constitute the starburst, the e-folding time of the exponential star formation history (SFH) and its age. The spheroidal model also has three parameters, which are the e-folding time of the delayed exponential SFH, the optical depth of the galaxy and the intensity of starlight. The dust and star distribution in the spheroidal galaxy is assumed to have a S\'ersic profile with $n=4$. The parameters of the starburst and spheroidal models are also listed in Table \ref{tab:parameters}.

\begin{table*}
	\centering
\caption{Parameters of the models used in this paper, symbols used, their assumed ranges and summary of other information about the models. The parameter that is fixed is shown in brackets. In the Fritz et al. (2006) model there are two additional parameters that define the density distribution in the radial direction ($\beta$) and azimuthal direction ($\gamma$). In this paper we assume $\beta=0$ and $\gamma=4$. In the SKIRTOR model there are two additional parameters that define the density distribution in the radial direction ($p$) and azimuthal direction ($q$). In this paper we assume $p=1$ and $q=1$. In addition, the SKIRTOR library fixes the fraction of mass inside clumps to 97 per cent. There are three additional scaling parameters for the starburst, spheroidal host and AGN torus models, $f_{SB}$, $f_{sph}$ and $f_{AGN}$, respectively. \\ \\}
	\label{tab:parameters}
 \resizebox{\textwidth}{!}{\begin{tabular}{llll} % four columns, alignment for each
 \hline
		Parameter &  Symbol & Range &  Comments\\
		\hline
                 &  &  & \\
{\bf CYGNUS Starburst}  &  &  & \\
                 &  &  \\
Initial optical depth of GMCs & $\tau_v$  &  50$-$250  &  \cite{efstathiou00}, \cite{efstathiou09} \\
Starburst SFR e-folding time       & $\tau_{*}$  & (20) Myr  & Incorporates \cite{bru93,bru03}  \\
Starburst age      & $t_{*}$   &  5$-$35 Myr &  Metallicity=solar, Salpeter IMF \\ 
                  &            &  & Standard galactic dust mixture with PAHs\\
                  &            &  &  \\                 
                   
{\bf CYGNUS Spheroidal Host}  &  &  &  \\
                 &  &  \\
Spheroidal SFR e-folding time      & $\tau^s$  &  0.125$-$8 Gyr  & \cite{efs03}, \cite{efs21}  \\
Starlight intensity      & $\psi^s$ &  1$-$17 &  Incorporates \cite{bru93,bru03} \\ 
Optical depth     & $\tau_{v}^s$ & 0.1$-$15 &  Range of metallicities, Salpeter IMF\\ 
                  &            &  & Standard galactic dust mixture with PAHs \\
                  &            &  &  \\               
                               
{\bf CYGNUS AGN torus}  &  &    &  \\
                 &  &  &  \\
Torus equatorial UV optical depth   & $\tau_{uv}$  &  260$-$1490 &  Smooth tapered discs\\  
Torus ratio of outer to inner radius & $r_2/r_1$ &  20$-$100 & \cite{efstathiou95}, \cite{efstathiou13} \\   
Torus half-opening angle  & $\theta_o$  &  30\degr$-$75\degr & Standard galactic dust mixture without PAHs\\ 
Torus inclination     & $\theta_i$  &  0\degr$-$90\degr &  The subranges $\theta_o$$-$90\degr \hspace{1pt} and 0\degr$-$$\theta_o$ are assumed for\\ 
                  &            &  &  AGN\_type=2 and AGN\_type=1, respectively. \\ 
                 &            & \\
{\bf \cite{fritz06} AGN torus}  &  &  &   \\
                 &  &  & \\
Torus equatorial optical depth at 9.7 $\mu m$  & $\tau_{9.7\mu m}$ &  1$-$10 & Smooth flared discs \\  
Torus ratio of outer to inner radius & $r_2/r_1$ &  10$-$150 &  \cite{fritz06}\\   
Torus half-opening angle  & $\theta_o$  &  20\degr$-$70\degr & Standard galactic dust mixture without PAHs\\ 
Torus inclination     &  $\theta_i$ &  0\degr$-$90\degr &  The subranges $\theta_o$$-$90\degr \hspace{1pt} and 0\degr$-$$\theta_o$ are assumed for\\ 
                  &            &  &  AGN\_type=2 and AGN\_type=1, respectively. \\    
                 &            & &  \\
{\bf SKIRTOR AGN torus}  &  &   &  \\
                 &  &  &  \\
Torus equatorial optical depth at 9.7 $\mu m$  &  $\tau_{9.7\mu m}$ &  3$-$11 & Two-phase flared discs \\  
Torus ratio of outer to inner radius & $r_2/r_1$ &  10$-$30 &  \cite{sta12,stal16} \\   
Torus half-opening angle  & $\theta_o$ &  20\degr$-$70\degr &  Standard galactic dust mixture without PAHs\\ 
Torus inclination     &  $\theta_i$ &  0\degr$-$90\degr & The subranges $\theta_o$$-$90\degr \hspace{1pt} and 0\degr$-$$\theta_o$ are assumed for\\ 
                  &            &  & AGN\_type=2 and AGN\_type=1, respectively. \\   
                 &            & &  \\
{\bf \cite{sieb15} AGN torus}  &  &   &  \\
                 &  &  &  \\
Cloud volume filling factor (per cent)   & $V_c$ &  1.5$-$77  & Two-phase anisotropic spheres \\  
Optical depth of the individual clouds & $A_c$  &  0$-$45 & \cite{sieb15}\\
Optical depth of the disc mid-plane & $A_d$  &  50$-$500 &  Fluffy dust mixture without PAHs\\ 
Inclination     &  $\theta_i$ &   0\degr$-$90\degr & The subranges 45\degr$-$90\degr \hspace{1pt} and 0\degr$-$45\degr are assumed for\\ 
                  &            &  &  AGN\_type=2 and AGN\_type=1, respectively. \\
                  &            &  &   \\                     
		\hline
	\end{tabular}}
\end{table*}

SMART can extract the fitted model parameters but also, using the routine \textit{post\_SMART} we have developed, a number of other physical quantities, which are listed in Table \ref{tab:derived}. For all the fits we fixed the e-folding time of the exponentially decaying SFR of the starburst at $\tau_{*} = 2 \times 10^7 yr$. We have a total of 12 parameters in the fits, including the scaling factors for each of the three components. Table \ref{tab:fitted} gives selected fitted parameters and their errors. A number of physical quantities we extracted are listed in Tables \ref{tab:extractedA} and \ref{tab:extractedB}.

The SFR of the starburst and the spheroidal component are computed self-consistently by the radiative transfer models, which incorporate the stellar population synthesis models of \cite{bru93,bru03}. We assume a Salpeter initial mass function (IMF) with a metallicity 6.5 per cent of solar for the spheroidal model and solar metallicity for the starburst model. In Table \ref{tab:extractedA} we list the SFR of the starburst averaged over the age of the starburst, $\dot{M}^{age}_{*}$, where the age is determined from the fit.

The anisotropy of the emission from the torus, which is a feature of all the torus models we consider in this work, requires a correction to the observed luminosity of AGN to obtain their intrinsic luminosity. \cite{efstathiou06}, \cite{efs22} and \cite{smart} defined the anisotropy correction factor $A$, which is essentially the factor by which we need to multiply the observed luminosity to get the true luminosity:
\begin{equation}
A(\theta_i) = {{\int_0^{\pi/2} ~~S(\theta_i') ~~sin \theta_i' ~~d \theta_i' } \over {S(\theta_i)}}~~,
\end{equation}
where $S(\theta_i)$ is the bolometric emission over the relevant wavelength range. $A(\theta_i)$ is generally different for the infrared and bolometric luminosities and is significant for all the AGN torus models used here. In Table \ref{tab:extractedB} we list the anisotropy correction factor $A$ predicted by the four different torus models.

\begin{figure*}
	\begin{center}
      {\includegraphics[width=70mm]{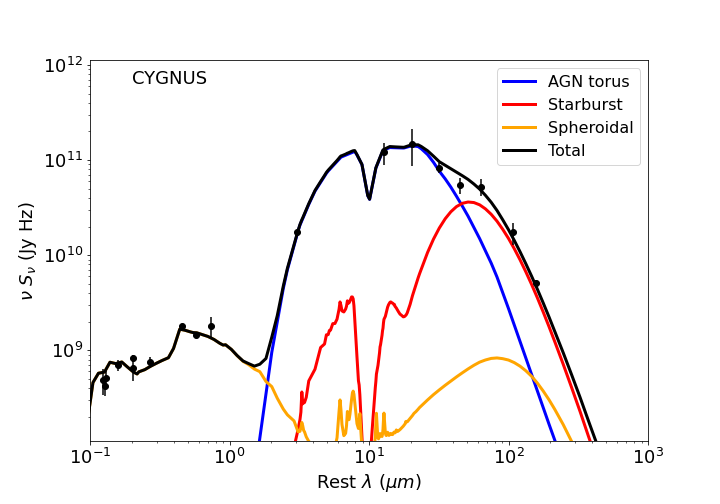}}    
      {\includegraphics[width=70mm]{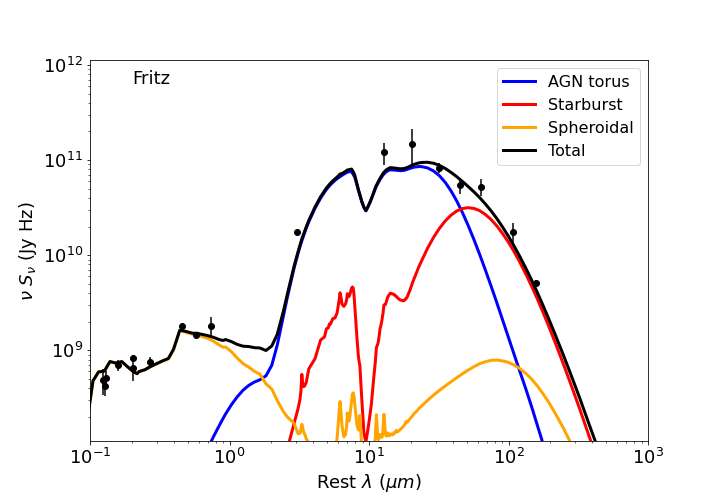}}    
      {\includegraphics[width=70mm]{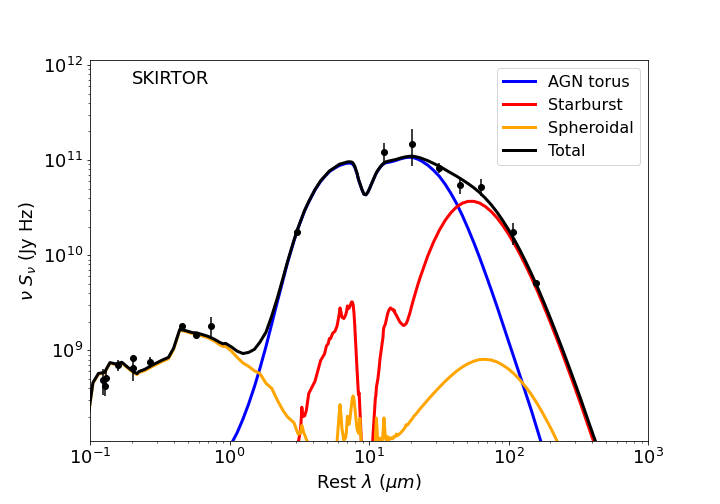}} 
      {\includegraphics[width=70mm]{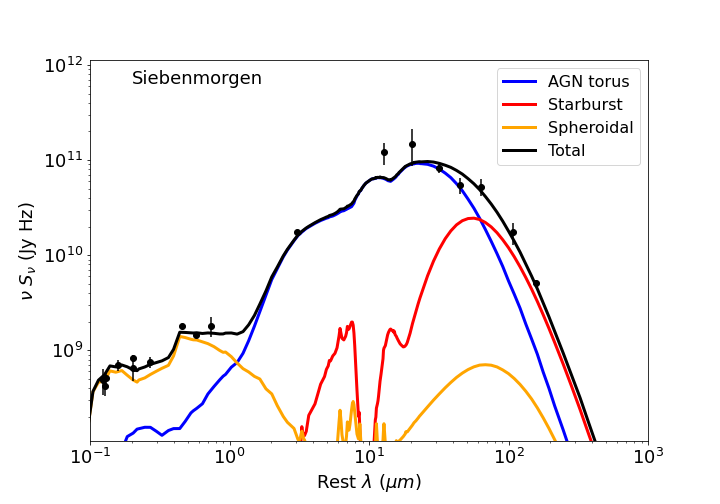}}
      {\includegraphics[width=70mm]{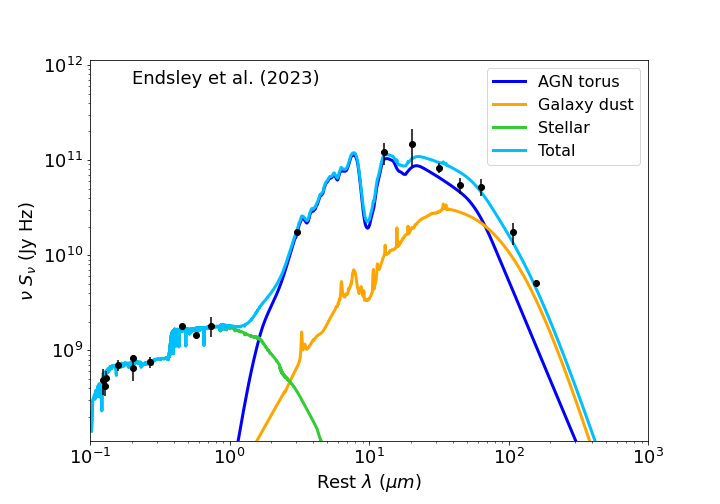}}
\caption{Comparison SED fit plots of COS-87529. The AGN torus, starburst, spheroidal host, galaxy dust, stellar and total emission are plotted as shown in the legend. The top left panel shows fits with the CYGNUS combination of models. The top right panel shows fits with the CYGNUS AGN torus model replaced by the Fritz et al. (2006) model, the middle left replaces the CYGNUS AGN torus model with the SKIRTOR model, while the middle right panel replaces the CYGNUS AGN torus model with the Siebenmorgen et al. (2015) model. In the bottom panel we plot in the same style the best-fitting models of Endsley et al. (2023).}
\label{fig:4AGN}      
      \end{center}
\end{figure*}

\begin{table}
	\centering
\caption{Derived physical quantities and the symbol used.}
	\label{tab:derived}
	\begin{tabular}{llll} % two columns, alignment for each
		\hline
		Physical quantity &  Symbol \\
		\hline  
Observed torus luminosity              & $L_{AGN}^{o}$    \\ 
Corrected torus luminosity             & $L_{AGN}^{c}$  \\ 
Starburst luminosity                   & $L_{SB}$  \\
Spheroidal host luminosity             & $L_{sph}$  \\
Total observed luminosity             & $L_{tot}^{o}$  \\ 
Total corrected luminosity             & $L_{tot}^{c}$  \\ 
Starburst SFR (averaged over its age)   & $\dot{M}_*^{age}$  \\
Spheroidal SFR                         & $\dot{M}_{sph}$  \\
Total SFR                              & $\dot{M}_{tot}$  \\
Spheroidal stellar mass                & $M^{*}_{sph}$  \\ 
Starburst stellar mass                 & $M^{*}_{SB}$  \\ 
Total stellar mass                     & ${M}^{*}_{tot}$  \\ 
AGN fraction                           & $F_{AGN}$  \\
Anisotropy correction factor           & $A$  \\ 
Corrected bolometric AGN luminosity    & $L_{bol,AGN}$  \\
Estimated black hole mass              & $M_{BH}$  \\
		\hline
	\end{tabular}
\end{table}

\section{Results and Discussion}\label{sec:results}

\begin{table*}
	\centering
\caption{Minimum reduced $\chi^2$ and selected fitted parameters for COS-87259. We also list for comparison the best-fitting results from Endsley et al. (2023).}
	\label{tab:fitted}
 \begin{tabular}{ lcccccccc }
 \hline 
AGN torus model &  $\chi^{2}_{min, \nu}$ & $\tau^s \ (10^7yr)$ & $\tau_v$ & $t_{*} \ (10^7yr)$  & & $\theta_o \ (\degr)$  & $\theta_i \ (\degr)$ \\ 
 \hline $  $ &  &  &  &  & $\tau_{9.7\mu m}$ \\  
 CYGNUS & 3.0  &    $16.2^{3.4}_{-3.0}$  &    $108.1^{15.6}_{-25.1}$ & $1.3^{0.3}_{-0.3}$ & $4.6^{0.3}_{-0.2}$  &    $48.8^{5.0}_{-6.5}$  &    $78.5^{6.8}_{-3.0}$\\
 \\   
 $  $ &  &  &  &  & $\tau_{9.7\mu m}$ \\ 
 \cite{fritz06} & 7.7  &    $17.5^{2.5}_{-2.7}$  &    $87.1^{6.6}_{-4.4}$  &  $1.8^{1.1}_{-1.0}$  &  $6.9^{2.5}_{-2.5}$  &    $25.1^{2.9}_{-2.2}$  &    $79.1^{6.3}_{-5.0}$\\ 
 \\   
 $  $ &  &  &  &  & $\tau_{9.7\mu m}$ \\ 
 SKIRTOR & 3.5  &        $15.3^{4.3}_{-1.8}$  &    $122.3^{73.2}_{-40.3}$  &  $1.6^{0.8}_{-0.5}$  &  $6.6^{1.3}_{-0.7}$  &    $33.6^{7.0}_{-8.3}$  &    $76.8^{5.4}_{-4.4}$\\  
 \\   
 $  $ &  &  &  &   &  $A_d$ \\ 
 \cite{sieb15} &   5.2  &    $14.6^{6.8}_{-1.3}$  &    $127.4^{63.8}_{-30.2}$  &  $2.3^{0.5}_{-0.5}$  &  $95.9^{85.9}_{-26.6}$  &    $0.0$  &    $59.3^{2.4}_{-1.1}$\\  
\\ \hdashline 
 $  $ &  &  &  &  & $\tau_{9.7\mu m}$ \\
 \cite{endsley23} & 6.1  &    &   &    &   $2.3^{0.1}_{-0.1}$  \\  \hline              
 \end{tabular}
\end{table*}

Our analysis allows us to assess which AGN torus model best fits the observational data for COS-87529. We find that the CYGNUS AGN torus model, with the tapered disc geometry, provides the best fit to the SED. In Table \ref{tab:fitted} we observe that there are significant differences in the minimum reduced $\chi^2$ ($\chi^{2}_{min, \nu}$), with the CYGNUS model giving the minimum value of $\chi^{2}_{min, \nu}=3$, the SKIRTOR model giving a $\chi^{2}_{min, \nu}=3.5$, the \cite{sieb15} model giving a $\chi^{2}_{min, \nu}=5.2$ and the \cite{fritz06} model giving the highest value of $\chi^{2}_{min, \nu}=7.7$. As in \cite{smart}, we assumed that an uncertainty of 15 per cent realistically represents our confidence in the model SEDs. We also assumed 15 per cent of uncertainty in the observed photometry. We used the same approach to calculate the minimum reduced $\chi^2$ for the best-fitting model of \cite{endsley23}, which is found to be $\chi^{2}_{min, \nu}=6.1$. This value is higher than that calculated for all model combinations used in this work, except from the combination with the \cite{fritz06} torus model. We note that in the results presented by \cite{endsley23} the 3.6 $\mu m$ band was not included. As the authors note, if the 3.6 $\mu m$ band is included, their fitting procedure gives a $\chi^{2}_{min, \nu}=11.8$. In our calculations for the minimum reduced $\chi^2$ of \cite{endsley23} we included the 3.6 $\mu m$ band, in order to be able to compare it with the values provided by our model combinations. The calculated minimum reduced $\chi^2$, as well as the SED fit plots in Fig. \ref{fig:4AGN}, show that the CYGNUS model is the most successful in fitting the rest frame mid-infrared data, where the SED peaks, and is therefore expected to give the highest AGN torus luminosity. The plots with the SKIRTOR model and with the models of \cite{endsley23} are also satisfactory, whereas the \cite{fritz06} and \cite{sieb15} models fail to reproduce the peak of the SED. All fits slightly underperdict the 5.8 $\mu m$ IRAC point, which, however, has a large uncertainty. The fit to this band could be improved by increasing the metallicity of the spheroidal component. Considering that the CYGNUS and SKIRTOR models fit considerably better than the \cite{fritz06} and \cite{sieb15} models, we consider on the results of only these two combinations in the rest of this discussion.

According to the best fits, the luminosity of this object is dominated by the AGN with an infrared ($1-1000~\mu m$) luminosity (including the anisotropy correction) of $5.5-10 \times 10^{13} L_\odot$. Our method predicts a significant anisotropy correction factor $A$ in the range $1.4-2.3$. The SKIRTOR model predicts the lowest $A$ and the CYGNUS model the highest. The CYGNUS model also predicts the highest AGN torus luminosity of $10^{14} L_\odot$. A high SFR is predicted by our method, ranging between $1985-2001~M_\odot yr^{-1}$. The estimate of SFR from \cite{endsley23} assumes a Kroupa IMF. We therefore divided by a factor of 0.66 to convert to a Salpeter IMF. The SFR estimate of \cite{endsley23} of 2439 $M_\odot yr^{-1}$ is significantly higher than any of the estimates from the combinations of models we explored.

The starburst model, which is used for all the fits presented in this work, assumes the starburst consists of an ensemble of giant molecular clouds (GMCs), which are at different evolutionary stages. Each GMC is assumed to have an initial radius of 50 pc and an initial optical depth in the \textit{V} band of $\tau_V$. The model does not make an assumption about the distribution of GMCs in space. From the luminosity of the starburst in COS-87259 we can infer the minimum volume that is needed to contain the GMCs that constitute the starburst. The following argument was first used by \cite{efstathiou09} to estimate the size of submillimetre galaxies. In the model of \cite{efstathiou00,efstathiou09} a molecular cloud illuminated by a 10 Myr old instantaneous burst has a luminosity of $\sim 3 \times 10^8 L_\odot$ and a radius of 50 pc. The minimum radius $R_{\rm min}$ of the sphere that is needed to contain a starburst of luminosity $L$ is therefore:
\begin{equation}
R_{\rm min} ~ \approx ~ 0.05 ~ (~{{L} \over {3 \times  10^8 L_\odot}~})^{1/3} ~~ {\rm kpc}~~.
\end{equation}
So, for a starburst with a luminosity of $8 \times 10^{12} L_\odot$, which is approximately the best-fitting value for the luminosity of the starburst in both the CYGNUS and SKIRTOR case, the minimum radius is $R_{\rm min} \sim 1.38$ kpc. This is much less than the physical size inferred from the ALMA dust continuum observations \citep{endsley23}. However, the distribution of GMCs can be more extended and compatible with the ALMA observations.

The stellar mass of the galaxy is predicted to be in the range $11.6-12.4 \times 10^{10} M_\odot$. The stellar mass estimate of $25.8 \times 10^{10} M_\odot$ by \cite{endsley23} is about a factor of two higher than the estimate from any of the combinations of models we explored. We also note that \cite{vanm23} estimate that the stellar mass is between $2.8-23.6 \times 10^{10} M_\odot$, depending on the SED fitting code used. A significant fraction of the stellar mass predicted by our method, ranging between $2.4-2.9 \times 10^{10} M_\odot$, formed in the current starburst episode in the galaxy. The total stellar mass formed in the current starburst is predicted to approximately double by the end of the episode. The AGN fraction is predicted to be in the range $86-92$ per cent. The method predicts a relatively young starburst age of $1.3-1.6 \times 10^7yr$. The torus half-opening angle is predicted to be in the range $33.6\degr-48.8\degr$, with the best-fitting CYGNUS model predicting the highest value. This implies that the torus in obscured quasars like COS-87259 has a large covering factor and these objects can be at least as numerous as unobscured quasars. We note that, of course, in the case of the SKIRTOR model, which is a two-phase model, the clumpiness of the medium must be taken into account to calculate the covering factor.

\begin{table*}
	\centering
\caption{Selected extracted physical quantities for COS-87259. For the total luminosity we give the anisotropy-corrected luminosity. A description of the physical quantities is given in Table \ref{tab:derived}. We also list for comparison the values derived from Endsley et al. (2023). The reported SMART luminosities are integrated over $1-1000~\mu m$, where those of Endsley et al. (2023) are integrated over $8-1000~\mu m$.}
	\label{tab:extractedA}
 \begin{tabular}{ lcccccccc }
 \hline 
AGN torus model & $L_{AGN}^o$ & $L_{AGN}^c$ & $L_{SB}$ & $L_{sph}$ & $L_{tot}^c$ & $\dot{M}^{age}_{*}$ & $\dot{M}_{sph}$ \\ 
     &  $10^{13} L_\odot$ &  $10^{13} L_\odot$ & $10^{12} L_\odot$ & $10^{11} L_\odot$ & $10^{13} L_\odot$ & $M_\odot yr^{-1}$ & $M_\odot yr^{-1}$ \\  
 \hline CYGNUS & $4.4^{0.7}_{-0.6}$  & $10.0^{0.7}_{-0.8}$  &    $8.4^{1.3}_{-1.4}$  &    $3.7^{2.1}_{-1.0}$  &    $10.9^{0.7}_{-0.9}$  &    $1967.0^{358.0}_{-318.2}$  &    $30.7^{25.9}_{-13.7}$ \\ 
 \\   
 \cite{fritz06} &  $3.1^{0.6}_{-0.2}$ &  $5.7^{0.6}_{-1.3}$  &    $7.4^{0.8}_{-0.4}$  &    $3.7^{0.7}_{-0.9}$  &    $6.4^{0.6}_{-1.2}$  &    $1919.0^{144.2}_{-35.8}$  &    $37.0^{11.9}_{-15.5}$ \\ 
 \\   
 SKIRTOR &   $3.6^{0.7}_{-0.4}$  &  $5.5^{1.1}_{-0.9}$  &    $8.1^{1.3}_{-1.0}$  &    $3.3^{1.0}_{-0.6}$  &    $6.5^{1.0}_{-1.0}$  &    $1957.0^{245.7}_{-211.5}$  &    $28.0^{22.3}_{-9.9}$ \\  
 \\   
 \cite{sieb15} &   $3.0^{0.5}_{-0.4}$ &  $3.7^{0.7}_{-0.7}$  &    $5.1^{1.8}_{-0.5}$  &    $3.4^{1.0}_{-0.7}$  &    $4.4^{0.6}_{-0.9}$  &    $1285.0^{357.6}_{-79.0}$  &    $25.9^{18.4}_{-9.3}$ \\  
 \\ \hdashline
 \\
 \cite{endsley23} &  $2.5^{0.2}_{-0.2}$   &      &   $9.1^{1.0}_{-1.0}$ \\
 \hline               
 \end{tabular}
\end{table*}

\begin{table*}
	\centering
\caption{Other extracted physical quantities for COS-87259. A description of the physical quantities is given in Table \ref{tab:derived}. Endsley et al. (2023) assume a Kroupa IMF, so, to compare with our estimates, we divided their SFR and stellar mass by a factor of 0.66.}
	\label{tab:extractedB}
 \begin{tabular}{ lcccccccc }
 \hline 
AGN torus model & $\dot{M}_{tot}$ & ${M}^{*}_{sph}$ & ${M}^{*}_{SB}$ & ${M}^{*}_{tot}$& $F_{AGN}$ & $A$ & $L_{bol,AGN}$ & $M_{BH}$ \\ 
     &     $M_\odot yr^{-1}$ & $10^{10} M_\odot$ & $10^{10} M_\odot$ & $10^{10} M_\odot$ & $   $ & $   $ & $10^{13} L_\odot$ & $10^{9} M_\odot$ \\  
 \hline CYGNUS &    $2001.0^{351.0}_{-308.7}$  &    $9.6^{1.3}_{-1.5}$  &    $2.4^{0.4}_{-0.5}$  &    $11.6^{2.0}_{-1.1}$  &     $0.92^{0.01}_{-0.02}$  &     $2.3^{0.4}_{-0.3}$ & $10.3^{0.5}_{-1.1}$  &    $3.2^{0.2}_{-0.3}$\\  
\\   
 \cite{fritz06} &    $1954.0^{137.0}_{-24.9}$  &    $8.5^{0.9}_{-0.8}$  &    $3.1^{1.9}_{-1.7}$  &    $11.6^{1.4}_{-1.1}$  &     $0.88^{0.01}_{-0.04}$  &     $1.7^{0.1}_{-0.2}$ & $6.0^{0.3}_{-1.6}$  &    $1.8^{0.1}_{-0.5}$\\ 
\\   
 SKIRTOR &    $1985.0^{287.0}_{-284.6}$  &    $9.2^{1.3}_{-1.3}$  &    $2.9^{1.3}_{-0.8}$  &    $12.4^{1.4}_{-1.7}$  &     $0.86^{0.04}_{-0.03}$  &     $1.4^{0.3}_{-0.2}$ & $5.8^{0.8}_{-1.1}$  &    $1.8^{0.3}_{-0.3}$\\
\\   
 \cite{sieb15} &    $1353.0^{305.5}_{-102.9}$  &    $8.7^{1.8}_{-2.3}$  &    $2.8^{0.7}_{-0.3}$  &    $11.7^{1.6}_{-2.5}$  &     $0.85^{0.05}_{-0.01}$  &     $1.2^{0.1}_{-0.1}$ & $3.9^{0.6}_{-0.8}$  &    $1.2^{0.2}_{-0.3}$\\
 \\ \hdashline
\\  
 \cite{endsley23} &  $2439.4^{287.9}_{-287.9}$   &      &      &  $25.8^{10.6}_{-10.6}$    &   $0.73^{0.08}_{-0.08}$   &  & $5.1^{0.5}_{-0.5}$ &  $1.6^{0.2}_{-0.2}$   \\ \hline                
 \end{tabular}
\end{table*}

COS-87259 appears to be similar to the galaxy HELP\_J100156.75+022344.7 at a photometric redshift of $z\sim4.3$, discovered by \cite{efs21} as part of the Herschel Extragalactic Legacy Project (HELP; \citealt{shi21}), also in the COSMOS field. We fitted HELP\_J100156.75+022344.7 with all four combinations of models considered in this paper and this is discussed further in \cite{smart}. The CYGNUS model provides the best fit to the data also in the case of HELP\_J100156.75+022344.7. We note that the CYGNUS smooth torus model was also found to give the best overall fits for the sample of local ultraluminous infrared galaxies studied by \cite{efs22}.

\cite{efs21} estimated a space density of $1.8 \times 10^{-8}$ Mpc$^{-3}$ for obscured quasars in the redshift range $4-5$, which is about a factor of two higher than the space density of coeval quasars of the same luminosity \citep{akiy18,wang19}. The space density of quasars of the same luminosity drops very steeply with increasing redshift and at $z=6.7$ is about a factor of 20 lower \citep{wang19}. As discussed by \cite{endsley23}, the discovery of another obscured hyperluminous quasar in the 1.7 deg$^2$ COSMOS field at $z=6.7$ is therefore difficult to explain. What is interesting to consider, though, is the space density of this type of objects in the redshift interval $z=4-7$, given the detection of two obscured hyperluminous quasars within the boundaries of this redshift interval. 

The comoving volume between $z=4$ and 7 for COSMOS, which we assume has an area of 1.7 deg$^2$, is $4.7 \times 10^7$ Mpc$^{-3}$. So, the space density of obscured hyperluminous quasars in this redshift range is $4.3 \times 10^{-8}$ Mpc$^{-3}$. This must be compared with the space density of quasars in the same redshift range, which is approximately $0.2 \times 10^{-8}$ Mpc$^{-3}$ according to \cite{wang19}. We can conclude from this rough calculation that the obscured quasars are about a factor of 20 more numerous than unobscured quasars at $z=4-7$, but not by three orders of magnitude as suggested by \cite{endsley23}. 

In Tables \ref{tab:extractedA} and \ref{tab:extractedB} we compare our key extracted physical quantities with the results of \cite{endsley23}. SMART predicts a lower SFR, lower stellar mass and higher AGN torus luminosity.  The estimate of \cite{endsley23} for the AGN torus luminosity does not include an anisotropy correction, which, according to our best-fitting model, is $\sim$ 130 per cent (i.e. the anisotropy correction factor $A=2.3$).

 \cite{endsley23} estimate a bolometric luminosity of $5.1 \times 10^{13} L_\odot $ and therefore estimate a black hole mass (assuming the AGN is accreting at the Eddington limit) of $1.6 \times 10^{9} M_\odot$. We estimate a bolometric luminosity about a factor of two higher  ($10^{14} L_\odot$), so the black hole mass is predicted to be twice as large ($3.2 \times 10^{9} M_\odot$). As discussed by \cite{far22}, a high fraction of luminous obscured AGNs in local galaxy mergers may be accreting at a super-Eddington rate at some point in their evolution, which is close to the time of the coalescence of the merging nuclei. If this is the case in COS-87259, then the predicted black hole mass may be much lower. This may also explain why these obscured AGNs are about a factor of 20 more numerous than coeval quasars of the same luminosity (see discussion above). Conversely, if the accretion in this quasar is not super-Eddington, then cosmological coupling of black holes, as discussed in \cite{far23}, may explain the appearance of such a massive supermassive black hole so early in the history of the Universe.

One of the early and puzzling results with \textit{JWST} was the discovery of a population of compact and red sources, known as Little Red Dots (LRDs). LRDs are estimated to be at redshifts $z>5$ (e.g. \citealt{akins23,labbe23a,labbe23b,matthee24,koko24,perez24}). LRDs show broad emission lines and it has been suggested that most of them may harbour an AGN. LRDs are much less luminous than COS-87259 and have black hole masses which are $2-4$ orders of magnitude higher than expected from the local $M_{\bullet}-M_{\star}$ relation. It is, nevertheless, interesting that the \textit{JWST} observations also suggest the presence of numerous red and potentially obscured AGNs at $z>5$.

\section{Conclusions}\label{sec:conclusion}

We explored the physical properties of the hyperluminous obscured quasar COS-87259 at $z=6.853$ with the SMART SED fitting code \citep{smart}, which allows fitting with four different AGN torus models, in addition to a starburst and a spheroidal host galaxy model. Our main conclusions can be summarized as follows:
\begin{enumerate}

\item
We find that the best fit is provided by the smooth CYGNUS tapered disc model. The SKIRTOR two-phase model also provides a good solution with a slightly worse minimum reduced $\chi^2$ than CYGNUS. The \cite{fritz06} and \cite{sieb15} models provide much worse fits. The fits with the CYGNUS and SKIRTOR models are also much better than the fits by \citealt{endsley23}.

\item
Our method predicts an AGN fraction that exceeds 85 per cent, but also an SFR that exceeds 1980 $M_\odot yr^{-1}$. According to our best-fitting model, the predicted bolometric AGN luminosity and therefore the estimated black hole mass of COS-87259 is $10^{14} L_\odot$ and $3.2 \times 10^{9} M_\odot$ respectively, about 50 per cent higher than that predicted by \cite{endsley23}.

\item
It is very puzzling that an object with so extreme properties was discovered in the 1.7 deg$^2$ COSMOS field. It would be very interesting to estimate the space density of such objects in larger fields with good infrared coverage, which will be surveyed by \textit{Euclid} and \textit{Roman} in the next few years. 

\end{enumerate}

\section*{Acknowledgements}

We would like to thank an anonymous referee for useful comments and suggestions. We also thank Ryan Endsley and Jianwei Lyu for providing their best-fitting models plotted in Fig. \ref{fig:4AGN}. CV and AE acknowledge support from the projects CYGNUS (contract number 4000126896) and CYGNUS+ (contract number 4000139319) funded by the European Space Agency.

\section*{Data Availability}

The data underlying this article are available in the article or are publicly available in the literature.

%%%%%%%%%%%%%%%%%%%%%%%%%%%%%%%%%%%%%%%%%%%%%%%%%%

%%%%%%%%%%%%%%%%% APPENDICES %%%%%%%%%%%%%%%%%%%%%

%%%%%%%%%%%%%%%%%%%%%%%%%%%%%%%%%%%%%%%%%%%%%%%%%%

% \appendix

% \section{Some extra material}

%%%%%%%%%%%%%%%%%%%%%%%%%%%%%%%%%%%%%%%%%%%%%%%%%%

% Don't change these lines
\bsp	% typesetting comment
\label{lastpage}
\end{document}